\documentclass{amsart}
\usepackage{amsthm, amssymb, amsfonts}
\normalsize
\theoremstyle{definition}

\newtheorem{Thm}{Theorem}

\newtheorem{ddefinition}[Thm]{Definition}

\begin{document}
\newcommand {\Ab}{\mathbf{A}}  % AA insted of Ab produces error msg.
\newcommand {\Bb}{\mathbf{B}}
\newcommand {\Cb}{\mathbf{C}}
\newcommand {\Pb}{\mathbf{P}}
\newcommand {\Rb}{\mathbb{R}}
\newcommand {\Ib}{\mathbb{I}}
\newcommand {\mb}{\mathbf{m}}
\newcommand {\op}{\mathbf{\oplus}} % op has spacing nicer than oplus
\newcommand {\om}{\mathbf{\ominus}} % om has spacing nicer than oplus
\newcommand {\gyrab}{\gyr[a,b]}
\newcommand {\gyrba}{\gyr[b,a]}
\newcommand {\gyr}{{\rm gyr}}
\newcommand {\gyrabb}{\gyr[\ab,\bb]}
\newcommand {\Aut}{{\textstyle Aut}}
\newcommand {\ccdot}{\mathbf{\cdot }}
\newcommand {\ab}{\mathbf{a}}
\newcommand {\bb}{\mathbf{b}}
\newcommand {\cb}{\mathbf{c}}
\newcommand {\ub}{\mathbf{u}}
\newcommand {\vb}{\mathbf{v}}
\newcommand {\xb}{\mathbf{x}}
\newcommand {\yb}{\mathbf{y}}
\newcommand {\zb}{\mathbf{z}}
\newcommand {\wb}{\mathbf{w}}
\newcommand {\bz}{\mathbf{0}}
\newcommand {\CC}{\mathbb{C}}
\newcommand {\DD}{\mathbb{D}}
\newcommand {\inn}{\hspace{-0.1cm}\in\hspace{-0.1cm}}
\newcommand {\Rtwo}{\Rb^2}
\newcommand {\Rstwou}{{\Rb}_{s=1}^2}
\newcommand {\Rstwo}{{\Rb}_{s}^2}
\newcommand {\Db}{\mathbb{D}}
\newcommand {\vi}{\mathbb{V}}
\newcommand {\vs}{\mathbb{V}_s}
\newcommand {\gyruvb}{\gyr[\ub,\vb]}
\newcommand {\subE}{\!\lower.1ex \hbox {\tiny E}}
\newcommand {\ope}{\op_{_{\subE}}\!\,}
%kkk
%%%%%%%%%%%%%%%%%%%%%%%%%%%%%%%%%%%%%%%%%%%%%%%%%%%%%%%%%%%%%%%%%%%%%%%%%%%%%
%    Copy of "paper052_frommobius.tex" from cdpapers begins here
%%%%%%%%%%%%%%%%%%%%%%%%%%%%%%%%%%%%%%%%%%%%%%%%%%%%%%%%%%%%%%%%%%%%%%%%%%%%%
%
%/ paper052_the_gyrogroup.tex
%
% TITLE
% TITLE             From M\"obius To Gyrogroups
% TITLE

\pagebreak
\pagenumbering{arabic}
\begin{center}
% TITLE
\huge{
From M\"obius To Gyrogroups
     }
\end{center}
%\vskip.2cm
\begin{center}
Abraham A. Ungar \\
Amer.~Math.~Monthly, Vol.~115(2), (2008), pp.~138--144.
%Department of Mathematics\\
%North Dakota State University\\
%Fargo, ND 58105, USA\\
%Email: abraham.ungar@ndsu.edu\\
% May 1, 2006\\[6pt]
\end{center}

%\begin{quotation}
%{\bf ABSTRACT}\phantom{OO}
%\end{quotation}

%SECTION 0
\section{Introduction}\label{sec0}

The evolution from M\"obius to gyrogroups began in
1988 \cite{parametrization}, and is still ongoing in \cite{mybook01,mybook02}.
%It is fascinating enough
%to be made part of the lore learned by all
%undergraduate and graduate mathematics students.
Gyrogroups, a natural generalization of groups, lay a
fruitful bridge between nonassociative algebra and hyperbolic geometry,
just as groups lay a fruitful bridge between associative algebra and
Euclidean geometry.
More than 150 years have passed since
the German mathematician August Ferdinand M\"obius (1790\,--\,1868)
first studied the transformations that now bear his name
\cite[p.~71]{wright02}.
Yet, the rich structure he thereby exposed is still far from
being exhausted, as
the evolution from M\"obius to gyrogroups demonstrates.

%SECTION 1
\section{From M\"obius}\label{sec1}

M\"{o}bius transformations of the complex open unit disc
$\DD= \{z\in \CC :\, |z|<1 \}$ of the complex plane $\CC$
are studied in most books on function theory of one complex variable.
According to \cite[p.~176]{needham97}, these are important for at least
two reasons: (i) they play a central role in non-Euclidean geometry
\cite{mybook02},
and (ii) they are the only automorphisms of the disc.
Ahlfors' book \cite{ahlfors73},
{\it Conformal Invariants: Topics in Geometric Function Theory},
begins with a presentation of the
{\it polar decomposition}
of the most general
M\"obius self-transformation
of the complex open unit disc $\DD$,
\begin{equation} \label{eq01}
z \mapsto e^{i\theta} \frac{a+z}{1+\overline{a}z}=e^{i\theta} (a\op z)
\,,
\end{equation}
where $a,z\inn\DD$ and $\theta\inn\Rb$
\cite[p.~211]{fisher99}
\cite[p.~185]{greenkrantz}
\cite[pp.~177\,--\,178]{needham97}.
We present the M\"obius polar decomposition \eqref{eq01}
in a form that suggests the {\it M\"obius addition}, $\op$,
defined by the equation
\begin{equation} \label{eq02}
a\op z=\frac{a+z}{1+\overline{a}z}
\,.
\end{equation}
Naturally, M\"obius subtraction, $\om$, is given by $a\om z = a\op (-z)$,
so that $z\om z=0$ and $\om z=0\om z=0\op(-z)=-z$.

Interestingly, M\"obius addition possesses the
{\it automorphic inverse property}
\begin{equation} \label{eqphic}
\om (a\op b) = \om a\om b
\end{equation}
and the {\it left cancellation law}
\begin{equation} \label{eq02a}
\om a\op(a\op z)=z
\end{equation}
for all $a,b,z\inn\DD$.

M\"obius addition is neither commutative nor associative. To make it
look like a commutative binary operation in the disc, we define the
{\it gyrations}
\begin{equation} \label{eq03}
\gyrab=\frac{a\op b}{b\op a}=\frac{1+a\overline{b}}{1+\overline{a}b}
\end{equation}
for $a,b\inn\DD$.
Being unimodular complex numbers, gyrations represent rotations of the disc
about its center. 

Moreover, gyrations respect M\"obius addition:
\begin{equation} \label{eq04}
\gyr[a,b](c\op d)=\gyr[a,b]c\op\gyr[a,b]d
\end{equation}
for all $a,b,c,d\inn\DD$. Also,
the inverse of a gyration $\gyr[a,b]$ is the gyration $\gyr[b,a]$:
\begin{equation} \label{eq05}
\gyr^{-1}[a,b] = \gyr[b,a]
\,.
\end{equation}
Hence, gyrations are special automorphisms of the M\"obius groupoid
$(\DD,\op)$.
We recall that a groupoid $(G,\op)$ is a nonempty set, $G$, with
a binary operation, $\op$, and that an automorphism of a
groupoid $(G,\op)$ is a bijective self map $f$ of $G$ that respects its
binary operation $\op$, that is, $f(a\op b)=f(a)\op f(b)$.
The set of all automorphisms of a groupoid $(G,\op)$ forms
a group, denoted $\Aut (G,\op)$.

To emphasize that gyrations are automorphisms, they are also called
{\it gyroautomorphisms}.
The automorphism $\phi(z)=-z$ of the disc is not a gyroautomorphism
of the disc since the equation
$(1+a\bar{b})/(1+\bar{a}b)=-1$ for $a,b\inn\DD$
has no solution in the disc. Consequently, the gyroautomorphisms of the disc
do not form a group under gyroautomorphism composition.

It follows from the gyration definition in \eqref{eq03}
that M\"obius addition obeys the {\it gyrocommutative law}:
\begin{equation} \label{eq06}
a\op b=\gyrab(b\op a)
\end{equation}
for all $a,b\inn\DD$. The M\"obius gyrocommutative law \eqref{eq06}
is not terribly surprising since it is generated by the gyration definition.
But, we are not finished!

``Coincidentally,'' the gyration $\gyrab$ that ``repairs''
in \eqref{eq06} the breakdown of
commutativity in M\"obius addition ``repairs'' the breakdown of
associativity as well, giving rise to the following
gyroassociative laws (left and right):
\begin{equation} \label{eq07}
 \begin{split}
a\op(b\op z)&=(a\op b)\op\gyrab z \,, \\
(a\op b)\op z&=a\op (b\op\gyrba z)
 \end{split}
 \end{equation}
for all $a,b,z\in\DD$.
Hence, gyrations measure simultaneously the extent to which M\"obius addition
deviates from both associativity and commutativity.

M\"obius gyrations are expressible in terms of M\"obius addition.
Indeed, solving the first identity in \eqref{eq07} for
$\gyr[a,b]z$ by means of the M\"obius left cancellation law
\eqref{eq02a} we have
\begin{equation} \label{eq07a}
\gyr[a,b]z = \om(a\op b) \op \{a\op(b\op z)\}
\,.
\end{equation}

M\"obius gyrations $\gyr[a,b]$, $a,b\inn\DD$,
thus endow the M\"obius disc groupoid $(\DD,\op)$ with
a group-like structure, naturally called a {\it gyrogroup}.
In fact, M\"obius gyrations do more than that. They
contribute their own rich structure to the gyrogroup.
Thus, for instance, they
possess the {\it loop properties} (left and right)
%%%%%%%%%%%%%%%%%%%%%%%%%%%%%%%%%%%%%%%%%%%%%%%%%%%%%%%%%%%%%%%%%%%%
\begin{equation} \label{eq08}
\begin{split}
\gyr[a\op b,b] &= \gyr[a,b] \,,\\
\gyr[a,b\op a] &= \gyr[a,b]\,,
\end{split}
\end{equation}
%%%%%%%%%%%%%%%%%%%%%%%%%%%%%%%%%%%%%%%%%%%%%%%%%%%%%%%%%%%%%%%%%%%%
and the nested gyration identity
\begin{equation} \label{eq09}
\gyr[b,\om\gyr[b,a]a] = \gyr[a,b] \,.
\end{equation}
More gyration identities, along with their applications in
hyperbolic geometry and Einstein's special relativity,
are presented in \cite{mybook01,mybook02}.
An important hyperbolic geometric interpretation of M\"obius gyrations
as triangle defects is presented in \cite{vermeer05}, and
an important special relativistic interpretation of gyrations
as {\it Thomas precessions} is presented in \cite{unleashing05}
and noted in \cite{ungarthomas06}.

Guided by the group-like structure of M\"obius addition, we define a
{\it gyrogroup} to be a group in which the associative law is replaced
by the gyroassociative laws (left and right), and a
{\it gyrocommutative gyrogroup} to be a commutative group in which the
commutative and associative laws are replaced by the
gyrocommutative and gyroassociative laws. Like M\"obius gyrations,
the gyrations in the gyrocommutative and gyroassociative laws of
an abstract gyrogroup must be automorphisms
of the gyrogroup.

Thus, the disc $\DD$ with its M\"obius addition forms a
{\it gyrocommutative gyrogroup} $(\DD,\op)$
where, strikingly, the concepts of gyrocommutativity and gyroassociativity
of the gyrogroup preserve the flavor of their classical counterparts
\cite{mybook01,mybook02}.
Mathematical coincidences are not accidental!
Hence, in particular, the gyrations that ``coincidentally'' repair in
\eqref{eq06}\,--\,\eqref{eq07} the breakdown of commutativity and associativity
need not be peculiar to M\"obius addition, as shown in
\cite{tuvalungar01,tuvalungar02}.
%
%Indeed, the generalization of groups into gyrogroups that M\"obius addition suggests
%bears an intriguing resemblance to the generalization of the rational numbers
%to the real ones. The beginner is initially surprised to discover an
%irrational number, like $\sqrt{2}$, but soon later
%he or she is likely to realize
%that there are more irrational numbers than rational ones.
%Similarly, the gyrogroup structure of M\"obius addition initially comes
%as a surprise.
%But, interested explorers may soon realize that there are indications
%that, in some sense, there are more non-group gyrogroups than groups
%\cite{tuvalungar01}.

%SECTION 2
\section{To Gyrogroups}\label{sec2}

Taking the key features of the M\"obius groupoid $(\DD,\op)$ as
axioms, we are now in a position to present the formal
gyrogroup definition.

% DEFINITION NUMBER -
\begin{ddefinition}\label{defroupx}
{\bf (Gyrogroups).}
{\it
A groupoid $(G , \op )$
is a gyrogroup if its binary operation satisfies the following axioms.
In $G$ there is at least one element, $0$, called a left identity, satisfying

\noindent
(G1) \hspace{1.2cm} $0 \op a=a$

\noindent
for all $a \in G$. There is an element $0 \in G$ satisfying axiom $(G1)$ such
that for each $a\in G$ there is an element $\om a\in G$, called a
left inverse of $a$, satisfying

\noindent
(G2) \hspace{1.2cm} $\om a \op a=0\,.$

\noindent
Moreover, for any $a,b,c\in G$ there exists a unique element $\gyr[a,b]c \in G$
such that the binary operation obeys the left gyroassociative law

\noindent
(G3) \hspace{1.2cm} $a\op(b\op c)=(a\op b)\op\gyrab c\,.$

\noindent
The map $\gyr[a,b]:G\to G$ given by $c\mapsto \gyr[a,b]c$
is an automorphism of the groupoid $(G,\op)$, that is,

\noindent
(G4) \hspace{1.2cm} $\gyrab\in\Aut (G,\op) \,,$

\noindent
and the automorphism $\gyr[a,b]$ of $G$ is called
the gyroautomorphism, or the gyration, of $G$ generated by $a,b \in G$.
The operator $\gyr : G\times G\rightarrow\Aut (G,\op)$ is called the
gyrator of $G$.
Finally, the gyroautomorphism $\gyr[a,b]$ generated by any $a,b \in G$
possesses the left loop property

\noindent
(G5) \hspace{1.2cm} $\gyrab=\gyr [a\op b,b] \,.$
}
\end{ddefinition}

The gyrogroup axioms ($G1$)\,--\,($G5$)
in Definition \ref{defroupx} are classified into three classes:
%%%%%%%%%%%%%%%%%%%%%%%%%%%%%%%%%%%%%%%%%%%%%%%%%%%%%%%%%%%%%%%%%%%%%%%%%%%%%
\begin{itemize}
\item[$(1)$]
The first pair of axioms, $(G1)$ and $(G2)$, is a reminiscent of the
group axioms.
\item[$(2)$]
The last pair of axioms, $(G4)$ and $(G5)$, presents the gyrator
axioms.
\item[$(3)$]
The middle axiom, $(G3)$, is a hybrid axiom linking the two pairs of
axioms in (1) and (2).
\end{itemize}
%%%%%%%%%%%%%%%%%%%%%%%%%%%%%%%%%%%%%%%%%%%%%%%%%%%%%%%%%%%%%%%%%%%%%%%%%%%%%

As in group theory, we use the notation
$a \om b = a \op (\om b)$
in gyrogroup theory as well.

In full analogy with groups, gyrogroups are classified into gyrocommutative and
non-gyrocommutative gyrogroups.

% DEFINITION NUMBER 4
\begin{ddefinition}\label{defgyrocomm}
{\bf (Gyrocommutative Gyrogroups).}
{\it
A gyrogroup $(G, \oplus )$ is gyrocommutative if
its binary operation obeys the gyrocommutative law

\noindent
(G6) \hspace{1.2cm} $a\oplus b=\gyrab(b\oplus a)$

\noindent
for all $a,b\in G$.
}
\end{ddefinition}

Clearly, a (commutative) group is a degenerate
(gyrocommutative) gyrogroup whose gyroautomorphisms are
all trivial. The algebraic structure of  gyrogroups
is, accordingly, richer than that of groups.
Thus, without losing the flavor of the group structure
we have generalized it into the gyrogroup structure to suit
the needs of M\"obius addition in the disc.
Fortunately, the gyrogroup structure is by no means restricted
to M\"obius addition in the disc.
Rather, it abounds in group theory as demonstrated, for instance,
in \cite{tuvalungar01} and \cite{tuvalungar02}, where finite
and infinite gyrogroups, both gyrocommutative and
non-gyrocommutative, are studied.
Some first gyrogroup theorems, some of which are analogous to
group theorems, are presented in \cite[Chap.~2]{mybook02}.

%SECTION 3
\section{M\"obius Gyrogroups: From the Disc To The Ball}\label{sec3}

The gyrocommutative gyrogroup structure in Definition \ref{defgyrocomm}
is tailor-made for M\"obius addition in the disc. However, it
suits M\"obius addition in the ball of any real inner
product space as well.

Let us identify complex numbers of the complex plane $\CC$ with
vectors of the Euclidean plane $\Rtwo$ in the usual way:
\begin{equation} \label{eq6100}
\CC \ni u = u_1+iu_2 = (u_1,u_2) = \ub \in \Rtwo \,.
\end{equation}
Then, the equations
\begin{equation} \label{eq6101}
\begin{array}{c}
\bar{u}v+u\bar{v} = 2\ub\ccdot\vb \,,
\\[8pt]
|u| = \|\ub\|
\end{array}
\end{equation}
give the inner product and the norm in $\Rtwo$,
so that M\"obius addition in the disc $\DD$ of $\CC$
becomes M\"obius addition in the disc
$\Rstwou=\{\vb\inn\Rtwo:\|\vb\|<s=1\}$ of $\Rtwo$. Indeed,
it follows from \eqref{eq6101} that
%%%%%%%%%%%%%%%%%%%%%%%%%%%%%%%%%%%%%%%%%%%%%%%%%%%%%%%%%%%%%%%%%%%%
\begin{equation} \label{eq6102}
\begin{split}
\DD \ni
u\op v &= \frac{u+v}{1+\bar{u}v}                      \\
&= \frac{(1+u\bar{v})(u+v)} {(1+\bar{u}v)(1+u\bar{v})}\\
&= \frac{(1+\bar{u}v+u\bar{v}+|v|^2)u+(1-|u|^2)v}
        {1+\bar{u}v+u\bar{v}+|u|^2|v|^2}              \\
&= \frac{(1+2\ub\ccdot\vb+\|\vb\|^2)\ub+(1-\|\ub\|^2)\vb}
         {1+2\ub\ccdot\vb+\|\ub\|^2\|\vb\|^2}         \\
&= \ub\op\vb \in \Rstwou
\end{split}
\end{equation}
%%%%%%%%%%%%%%%%%%%%%%%%%%%%%%%%%%%%%%%%%%%%%%%%%%%%%%%%%%%%%%%%%%%%
for all $u,v\in\Db$ and all $\ub,\vb\in\Rstwou$.
The last equation in \eqref{eq6102} is a vector equation, so that
its restriction to the ball of the Euclidean two-dimensional space
is a mere artifact.
As such, it survives unimpaired in higher dimensions,
suggesting the following
definition of M\"obius addition in the ball of any real inner
product space.

%%%%%%%%%%%%%%%%%%%%%%%%%%%%%%%%%%%%%%%%%%%%%%%%%%%%%%%%%%%%%%%%%%%%
% DEFINITION NUMBER 3
\begin{ddefinition}\label{defmobiusadd}
{\bf (M\"obius Addition in the Ball).}
Let $\vi$ be a real inner product space \cite{marsden74}, and let
$\vs$ be the $s$-ball of $\vi$,
\begin{equation} \label{eqsball}
\vs = \{\vs\in\vi : \|\vb\|<s\}
\,,
\end{equation}
for any fixed $s>0$.
M\"obius addition $\op$ is a binary operation in $\vs$ given by
the equation
%%%%%%%%%%%%%%%%%%%%%%%%%%%%%%%%%%%%%%%%%%%%%%%%%%%%%%%%%%%%%%%%%%%%
 \begin{equation} \label{eq696}
\ub\op\vb
 = \frac{(1+\frac{2}{s^2}\ub\ccdot\vb+\frac{1}{s^2}\|\vb\|^2 )\ub
        +(1-\frac{1}{s^2}\|\ub\|^2)\vb}
        {1+\frac{2}{s^2}\ub\ccdot\vb+\frac{1}{s^4}\|\ub\|^2\|\vb\|^2}
\,,
 \end{equation}
where $\ccdot$ and $\|\ccdot\|$ are the inner product and norm that the
ball $\vs$ inherits from its space $\vi$ and where, ambiguously,
+ denotes both addition of real numbers on the real line and addition
of vectors in $\vi$.
\end{ddefinition}
%%%%%%%%%%%%%%%%%%%%%%%%%%%%%%%%%%%%%%%%%%%%%%%%%%%%%%%%%%%%%%%%%%%%

Without loss of generality, one may select $s=1$ in
Definition \ref{defmobiusadd}.
We, however, prefer to keep $s$ as a free positive parameter in order
to exhibit the result that in the limit as $s\rightarrow\infty$,
the ball $\vs$ expands to the whole of its real inner product space $\vi$,
and M\"obius addition $\op$ reduces to vector addition in $\vi$.

Being special automorphisms,
M\"obius gyrations form an important ingredient of the
M\"obius disc gyrogroup $(\DD,\op)$.
Following the extension of M\"obius addition from the disc to the ball,
it is desirable to extend M\"obius gyrations from the disc to the ball
as well. On first glance this task seems impossible, since
M\"obius gyrations in the disc are given by \eqref{eq03} in terms
of a division of complex numbers by nonzero complex numbers, an operation
that cannot be extended from the disc to the ball.
Fortunately, however, identity \eqref{eq07a} comes to the rescue.
This identity
expresses gyrations in terms of M\"obius addition in the disc, an
operation that we have just extended from the disc to the ball
in \eqref{eq6102}, as formalized in Definition \ref{defmobiusadd}.

As suggested by \eqref{eq07a}, and in agreement with
Definition \ref{defroupx} of gyrogroups,
the definition of M\"obius gyrations in the ball follows.
%%%%%%%%%%%%%%%%%%%%%%%%%%%%%%%%%%%%%%%%%%%%%%%%%%%%%%%%%%%%%%%%%%%%
% DEFINITON NUMBER 4
\begin{ddefinition}\label{defgyr} 
{\bf (M\"obius Gyrations in the Ball).}
Let $(\vs,\op)$ be a M\"obius groupoid. The gyrator gyr is the map
$gyr:\vs\times\vs ~\rightarrow~\Aut(\vs,\op)$
given by the equation
\begin{equation} \label{rnjsd01}
\gyr[\ab,\bb]\zb = \om(\ab\op\bb)\op\{\ab\op(\bb\op\zb)\}
\end{equation}
for all $\ab,\bb,\zb\in\vs$.
The automorphisms $\gyr[\ab,\bb]$ of $\vs$ are
called gyrations, or gyroautomorphisms, of the ball $\vs$.
\end{ddefinition}
%%%%%%%%%%%%%%%%%%%%%%%%%%%%%%%%%%%%%%%%%%%%%%%%%%%%%%%%%%%%%%%%%%%%

It is anticipated in Definition \ref{defgyr} that gyrations of the ball
are automorphisms of the ball groupoid $(\vs,\op)$.
To show that this is indeed the case, we note that \eqref{rnjsd01} can
be manipulated by means of Definition \ref{defmobiusadd} of M\"obius addition,
obtaining (with the help of computer algebra)
\begin{equation} \label{hdge1}
\begin{split}
\gyruvb\wb &= \om(\ub\op\vb)\op\{\ub\op(\vb\op\wb)\} \\[8pt]
&= \wb + 2\frac{A\ub+B\vb}{D} \,,
\end{split}
\end{equation}
where
%%%%%%%%%%%%%%%%%%%%%%%%%%%%%%%%%%%%%%%%%%%%%%%%%%%%%%%%%%%%%%%%%%%%
\begin{equation} \label{hdgej2}
\begin{split}
A &=-\frac{1}{s^4}\ub\ccdot\wb\|\vb\|^2 + \frac{1}{s^2}\vb\ccdot\wb
    +\frac{2}{s^4}(\ub\ccdot\vb)(\vb\ccdot\wb) \,, \\[8pt]
B &=-\frac{1}{s^4}\vb\ccdot\wb\|\ub\|^2 - \frac{1}{s^2}\ub\ccdot\wb \,, \\[8pt]
D &= 1 + \frac{2}{s^2}\ub\ccdot\vb + \frac{1}{s^4}\|\ub\|^2 \|\vb\|^2
\end{split}
\end{equation}
% MATHEMATICA "mobius001",  MATLAB "gyrmb.m"
%%%%%%%%%%%%%%%%%%%%%%%%%%%%%%%%%%%%%%%%%%%%%%%%%%%%%%%%%%%%%%%%%%%% 
for all $\ub,\vb,\wb\in\vs$.
Owing to the Cauchy-Schwarz inequality \cite[p.~20]{marsden74},
$D>0$ for $\ub$ and $\vb$ in the ball $\vs$.
By expanding the domain of $\wb$ from the ball $\vs$ to the space $\vi$
in \eqref{hdge1}\,--\,\eqref{hdgej2},
we can extend the gyrations $\gyr[\ub,\vb]$
to linear maps of $\vi$ for all $\ub,\vb\in\vs$.

It follows from
\eqref{hdge1}\,--\,\eqref{hdgej2}
straightforwardly (the use of computer
algebra is recommended) that
\begin{equation} \label{eq1ffmzn}
\gyr[\vb,\ub](\gyr[\ub,\vb]\wb) = \wb
\end{equation}
for all $\ub,\vb,\wb\inn\vs$, so that gyrations of the ball are invertible,
the inverse of $\gyr[\ub,\vb]$ being $\gyr[\vb,\ub]$.

Furthermore,
it follows from \eqref{hdge1}\,--\,\eqref{hdgej2}
straightforwardly (the use of computer
algebra is recommended) that M\"obius gyrations of the ball
preserve the inner product that the ball $\vs$ inherits from
its real inner product space $\vi$, that is,
\begin{equation} \label{eq1ffmob03}
\gyruvb\ab\ccdot\gyruvb\bb=\ab\ccdot\bb
\end{equation}
for all $\ab,\bb,\ub,\vb\in\vs$.
This implies, in turn, that M\"obius gyrations keep invariant the
norm that the ball $\vs$ inherits from
its real inner product space $\vi$. As such, M\"obius gyrations of
the ball preserve M\"obius addition in the ball, so that
M\"obius gyrations are automorphisms of the M\"obius ball groupoid
$(\vs,\op)$.
Being special automorphisms, M\"obius gyrations are also called
gyroautomorphisms of the ball.
The automorphism $\phi(\zb)=-\zb$ of the ball is not a gyroautomorphism
of the ball, as verified in \cite[p.~70]{mybook02}.
Hence, the gyroautomorphisms of the ball do not form a group
under gyroautomorphism composition.

Not unexpectedly, M\"obius addition $\op$ in the ball $\vs$ of any
real inner product space $\vi$ preserves the structure it has in the disc.
It thus gives rise to the M\"obius gyrocommutative gyrogroup $(\vs,\op)$
in the ball, studied in \cite{mybook02}.
M\"obius addition in the ball is known in the literature
as a {\it hyperbolic translation} \cite{ahlfors81,ratcliffe94}.
However, its gyrogroup structure went unnoticed until it was uncovered
in 1988 \cite{parametrization,ahlfors} in the context of
Einstein's special theory of relativity. Indeed, like M\"obius addition,
Einstein's velocity addition law of special relativity
gives rise to the Einstein gyrocommutative gyrogroup $(\vs,\ope)$
in the ball, studied in \cite{parametrization,mybook01,unleashing05}.

The evolution of M\"obius addition and gyrations in the disc
does not stop at the level
of gyrogroups. It continues into the regime of {\it gyrovector spaces}.
Strikingly, gyrovector spaces form the setting for hyperbolic
geometry just as vector spaces form the setting for
Euclidean geometry. In particular, M\"obius gyrovector spaces form the
setting for the Poincar\'e ball model of hyperbolic geometry while,
similarly, Einstein gyrovector spaces form the setting for the Beltrami-Klein
ball model of hyperbolic geometry
\cite{mybook01,mybook02,ungardiff05,gyroparallelogram06,walterrev2002}.
Accordingly, a gyrovector space approach to analytic hyperbolic geometry,
fully analogous to the common vector space approach to
Euclidean geometry \cite{hausner98}, is developed in \cite{mybook02}.
%
%Equations \eqref{eq01}\,--\,\eqref{eq09} of M\"obius addition and gyrations
%in the disc remain valid in the ball as well. Hence, they
%clearly pass the tests of simplicity, beauty, and most importantly,
%generality. Accordingly,
%the far-reaching evolution of M\"obius addition and gyrations in the disc
%into the abstract gyrogroup is likely to set new standards in
%twenty-first century hyperbolic geometry, special relativity theory,
%and nonassociative algebra.
%The symbiosis between these three theories vastly enriches them,
%largely due to the presence of gyrations.

%%%%%%%%%%%%%%%%%%%%%%%%%%%%%%%%%%%%%%%%%%%%%%%%%%%%%%%%%%%%%%%%%%%%%%%%%%%%%
%    Copy of "paper052_frommobius.tex" from cdpapers ends here
%%%%%%%%%%%%%%%%%%%%%%%%%%%%%%%%%%%%%%%%%%%%%%%%%%%%%%%%%%%%%%%%%%%%%%%%%%%%%

\bibliographystyle{amsplain}

\begin{thebibliography}{1}

\bibitem{ahlfors73}
Lars~V. Ahlfors.
\newblock {\em Conformal invariants: topics in geometric function theory}.
\newblock McGraw-Hill Book Co., New York, 1973.
\newblock McGraw-Hill Series in Higher Mathematics.

\bibitem{ahlfors81}
Lars~V. Ahlfors.
\newblock {\em M\"obius transformations in several dimensions}.
\newblock University of Minnesota School of Mathematics, Minneapolis, Minn.
  1981.

\bibitem{fisher99}
Stephen~D. Fisher.
\newblock {\em Complex variables}.
\newblock Dover Publications Inc. Mineola, NY, 1999.
\newblock Corrected reprint of the second (1990) edition.

\bibitem{tuvalungar01}
Tuval Foguel and Abraham~A. Ungar.
\newblock Involutory decomposition of groups into twisted subgroups and
  subgroups.
\newblock {\em J. Group Theory}, {\bf 3} (2000) 27--46.

\bibitem{tuvalungar02}
Tuval Foguel and Abraham~A. Ungar.
\newblock Gyrogroups and the decomposition of groups into twisted subgroups and
  subgroups.
\newblock {\em Pac. J. Math.} {\bf 197} (2001) 1--11.

\bibitem{greenkrantz}
Robert~E. Greene and Steven~G. Krantz.
\newblock {\em Function theory of one complex variable}.
\newblock John Wiley \& Sons Inc. New York, 1997.

\bibitem{hausner98}
Melvin Hausner.
\newblock {\em A vector space approach to geometry}.
\newblock Dover Publications Inc., Mineola, NY, 1998.
\newblock Reprint of the 1965 original.

\bibitem{marsden74}
Jerrold~E. Marsden.
\newblock {\em Elementary classical analysis}.
\newblock W. H. Freeman and Co., San Francisco, 1974.
\newblock With the assistance of Michael Buchner, Amy Erickson, Adam
  Hausknecht, Dennis Heifetz, Janet Macrae and William Wilson, and with
  contributions by Paul Chernoff, Istv\'an F\'ary and Robert Gulliver.

\bibitem{wright02}
David Mumford, Caroline Series, and David Wright.
\newblock {\em Indra's pearls: The vision of Felix Klein}.
\newblock Cambridge University Press, New York, 2002.

\bibitem{needham97}
Tristan Needham.
\newblock {\em Visual complex analysis}.
\newblock The Clarendon Press Oxford University Press, New York, 1997.

\bibitem{ratcliffe94}
John~G. Ratcliffe.
\newblock {\em Foundations of hyperbolic manifolds}, volume 149 of {\em
  Graduate Texts in Mathematics}.
\newblock Springer-Verlag, New York, 1994.

\bibitem{parametrization}
Abraham~A. Ungar.
\newblock Thomas rotation and the parametrization of the {L}orentz
  transformation group.
\newblock {\em Found. Phys. Lett.} {\bf 1} (1988) 57--89.

\bibitem{ahlfors}
Abraham~A. Ungar.
\newblock Mobius transformations of the ball, {A}hlfors' rotation and
  gyrovector spaces.
\newblock In {\em Themistocles M. Rassias (ed.): Nonlinear analysis in geometry
  and topology}, pages 241--287. Hadronic Press, Palm Harbor, FL, 2000.

\bibitem{mybook01}
Abraham~A. Ungar.
\newblock {\em Beyond the {E}instein addition law and its gyroscopic {T}homas
  precession: The theory of gyrogroups and gyrovector spaces}, volume 117 of
  {\em Fundamental Theories of Physics}.
\newblock Kluwer Academic Publishers Group, Dordrecht, 2001.

\bibitem{mybook02}
Abraham~A. Ungar.
\newblock {\em Analytic hyperbolic geometry: Mathematical foundations and
  applications}.
\newblock World Scientific Publishing Co. Pte. Ltd., Hackensack, NJ, 2005.

\bibitem{unleashing05}
Abraham~A. Ungar.
\newblock Einstein's special relativity: Unleashing the power of its hyperbolic
  geometry.
\newblock {\em Comput. Math. Appl.} {\bf 49} (2005) 187--221.

\bibitem{ungardiff05}
Abraham~A. Ungar.
\newblock Gyrovector spaces and their differential geometry.
\newblock {\em Nonlinear Funct. Anal. Appl.} {\bf 10} (2005) 791--834.

\bibitem{gyroparallelogram06}
Abraham~A. Ungar.
\newblock The relativistic hyperbolic parallelogram law.
\newblock In {\em Geometry, integrability and quantization}. Softex, Sofia,
  (2006) 249--264.

\bibitem{ungarthomas06}
Abraham~A. Ungar.
\newblock Thomas precession: a kinematic effect of the algebra of {E}instein's
  velocity addition law. {C}omments on `deriving relativistic momentum and
  energy: {II}. three dimensional case'.
\newblock {\em European J. Phys.} {\bf 27} (2006) L17--L20.

\bibitem{vermeer05}
J.~Vermeer.
\newblock A geometric interpretation of {U}ngar's addition and of gyration in
  the hyperbolic plane.
\newblock {\em Topology Appl.} {\bf 152} (2005) 226--242.

\bibitem{walterrev2002}
Scott Walter.
\newblock Book {R}eview: {\it {B}eyond the {E}instein {A}ddition {L}aw and its
  {G}yroscopic {T}homas {P}recession: {T}he {T}heory of {G}yrogroups and
  {G}yrovector {S}paces}, by {A}braham {A}. {U}ngar.
\newblock {\em Found. Phys.} {\bf 32} (2002) 327--330.

\end{thebibliography}

\end{document}